\documentstyle[art10]{article}

\newcommand{\beq}{\begin{equation}}
\newcommand{\eeq}{\end{equation}}
\newcommand{\ep}{\varepsilon}
\newcommand{\vf}{\varphi}

\begin{document}

\title{
Integrable dynamics of a discrete curve \\
and the Ablowitz-Ladik hierarchy
\thanks{This work was supported by the 1993 agreement between Rome and Warsaw
Universities; by the grant 2-0168-91-01 KBN and by the INFN. }}

\author{ Adam Doliwa  \\
Institute of Theoretical Physics, Warsaw University\\
ul. Ho\.{z}a 69, 00-681 Warsaw, Poland
\and
Paolo Maria Santini \\
Dipartimento di Fisica, Universit\`{a} di Roma ``La Sapienza'' \\
and INFN, Sezione di Roma \\
P.le Aldo Moro 2, I-00185 Roma, Italy}

\date{}
\maketitle

\begin{abstract}
\noindent We show that the following elementary geometric properties of the
motion of a discrete (i.e. piecewise linear) curve select the integrable
dynamics of the Ablowitz-Ladik
hierarchy of evolution equations: i) the set of points describing the
discrete curve lie in the sphere $S^{3}$; ii) the distance between any two
subsequent points does not vary in time; iii) the dynamics does not depend
explicitly on the radius of the sphere. These results generalize to a discrete
context our previous work on continuous curves \cite{DolSa}.

\end{abstract}

%1993 PACS: 02.30.Jr, 02.40.Hw.

\bigskip
\bigskip
\bigskip
\bigskip
\bigskip

%\hbox to 5truein{ \hfil {\large {\bf Preprint IFT UW/6/94}}}

\bigskip
\bigskip
\bigskip
\bigskip
\bigskip
\bigskip
\bigskip
\bigskip

\pagebreak
\section{Integrable discrete dynamics and geometry}
\label{sec:IDG}
During the last two decades many important classes of nonlinear equations have
been solved. These integrable nonlinear equations appear in very different
forms: they are ordinary differential equations (ODE's)
(autonomous and not)\cite{D_P}, partial
differential equations (PDE's) \cite{ZMNP-AS}, algebraic and functional
equations \cite{DFS}. It was also possible to construct integrable discrete
analogues of all the important nonlinear ODE's and PDE's
\cite{Hir}\cite{Le}\cite{ZMNP-AS}\cite{Ni}\cite{FIK}\cite{Sh}; in
particular, Ablowitz and Ladik (AL) \cite{AL} have constructed an integrable
discrete analogue of the AKNS hierarchy \cite{AKNS}, associated with the
following discrete eigenvalue problem
\beq
\label{eq:dsp}
\left[ \begin{array}{c} \phi^{(1)}_{k+1} \\ \phi^{(2)}_{k+1} \end{array}
\right] = \left[ \begin{array}{cc} \zeta & q_{k} \\
  r_{k} & \zeta^{-1} \end{array} \right]
\left[ \begin{array}{c} \phi^{(1)}_{k} \\ \phi^{(2)}_{k} \end{array} \right]
\; \; ,
\eeq
where the complex fields $q_{k}$ and $r_{k}$ depend on the discrete
variable $k$ ($k \in {\bf Z}$), and the complex parameter $\zeta$ plays the
role of spectral parameter.

The interesting connections between the differential geometry of
sub\-ma\-ni\-folds
and integrable nonlinear PDE's, which have been shown many times throught the
years (the literature on this subject is very large and we refer to our
previous paper \cite{DolSa} for a historical review), can also be found at a
discrete level. For instance, discrete pseudospherical surfaces and the
discrete analogues of constant mean curvature surfaces are described by
integrable discrete analogues of the sine-Gordon (SG) \cite{MePin} and
sinh-Gordon
equations \cite{Pin}. Such discretizations were found by adapting the main
geometric properties of the continuous surfaces to a discrete level.

\medskip
\noindent In a recent paper \cite{DolSa} we have shown that the following three
elementary geometric properties of the motion of a curve select, among
all possible dynamics, {\bf integrable dynamics}:

\medskip

\noindent {\it Property 1.} The motion of the curve takes place in
 the $N$-dimensional sphere of radius $R$, denoted
by $S^{N}(R)$, $N>1$.

\medskip
\noindent {\it Property 2.} The curve does not stretch during the motion.

\medskip
\noindent {\it Property 3.} The dynamics of the curve does not depend
{\bf explicitly} on the radius $R$.

\medskip
\noindent These dynamics are described by integrable commuting nonlinear
evolution
equations in 1+1 (one spatial and one temporal) dimensions for the geodesic
curvatures of the curve. More precisely, for $N=2$ we have obtained the
modified Korteweg -- de Vries (mKdV)
hierarchy for the curvature $\kappa$. For $N=3$ we have obtained a hierarchy
of evolution equations for the geodesic curvature $\kappa=\kappa_{1}$ and
torsion $\tau=\kappa_{2}$, which can be transformed, through the Hasimoto
transformation, into the nonlinear Schr\"{o}dinger (NLS)
hierarchy. For $N>3$ we have obtained integrable
multicomponent generalizations of the above hierarchies. Also integrable
evolution equations "with sources" (with singular dispersion relation) have
been selected by {\it Properties 1-3}.

The goal of this paper is to construct large classes of geometrically
meaningful discrete evolution equations applying the above {\it Properties 1-3}
to the case of a discrete curve, where by discrete curve we mean just a
sequence of points. The obvious discrete generalization of {\it Properties 1-3}
is the following.

\medskip
\noindent {\it Property 1}'. The motion of the discrete curve takes place in
the $N$-dimensional sphere  $S^{N}(R)$ ($N>1$) of radius $R$.

\medskip
\noindent {\it Property 2}'. The distance between any two subsequent points
does not vary in time.

\medskip
\noindent {\it Property 3}'. The dynamics of the discrete curve does not depend
{\bf explicitly} on the radius $R$.

\medskip
\noindent We will consider the cases $N=2$ and $N=3$, showing that {\it
Properties 1'-3'} generate the AL hierarchy of evolution equations. In
particular we will prove that {\it Property 1}'  for $N=2$ and $N=3$ is
equivalent to the spectral problem (\ref{eq:dsp}) in the reductions
$r_{k}=-q_{k} \in {\bf R}$ and $r_{k}=-\bar{q}_{k}$ respectively.

In this paper we concentrate our attention on  continuous evolutions, i.e. time
$t$ is a continuous variable; we postpone to a subsequent work the study of
integrable discrete dynamics of a curve.

\section{The motion of a discrete curve on $S^{2}(R)$}
\label{sec:dis-S2}
The problem of a curve on $S^{2}$ is a special of case of that of a curve in
$S^{3}$,
but because of its pedagogical importance we have decided to treat it first.

\subsection{The discrete curve on $S^{2}(R)$}
Generalizing the problem of a
discrete curve on the plane (see, eg. \cite{Aminov}), we derive now the set of
equations which describe a discrete curve on $S^{2}(R)$. We indicate by ${\bf
\vec{r}}_{k}$, $(k\in{\bf Z})$ the position vector of the $k^{\rm th}$ point of
the sequence pointing from the centre of the sphere. The unit vector in the
direction of ${\bf \vec{r}}_{k}$  is denoted by ${\bf \hat{r}}_{k}$. The
oriented plane $\pi_{k}$ is given by the ordered pair $({\bf \vec{r}}_{k} \, ,
\, {\bf \vec{r}}_{k+1})$. By ${\bf \hat{t}}_{k}$ $({\bf \hat{t}}'_{k})$ we
denote the unit vector of the plane $\pi_{k}$ orthogonal to ${\bf \hat{r}}_{k}$
$({\bf \hat{r}}_{k+1})$ with the direction given by the orientation of the
plane $\pi_{k}$. Both ${\bf \hat{t}}_{k}$ and ${\bf \hat{t}}'_{k}$ are tangent
(in the points ${\bf \vec{r}}_{k}$ and ${\bf \vec{r}}_{k+1}$ correspondingly)
to the big circle passing through the points ${\bf \vec{r}}_{k}$ and ${\bf
\vec{r}}_{k+1}$.

In the following we assume that the segments of geodesics which join subsequent
points have the same length $\Delta$ for each pair of the sequence. The angle
$\alpha$ is one-half of the angle $\lambda\Delta$ (where $\lambda = 1 /R$ )
between the vectors ${\bf
\vec{r}}_{k}$ and ${\bf \vec{r}}_{k+1}$:
\beq
\alpha = \Delta/(2R) \; \; .
\eeq

%\epsffile{01.ps}

\bigskip

\bigskip

\noindent Together with the basis $({\bf \hat{r}}_{k},{\bf \hat{t}}_{k})$ of
the plane
$\pi_{k}$ we will use also the new basis $({\bf \hat{r}}'_{k},{\bf
\hat{a}}_{k})$ obtained from the old one by rotation of the angle $\alpha$
\beq \begin{array}{ccc}
{\bf \hat{r}}'_{k} & = & {\bf \hat{r}}_{k} \cos\alpha - {\bf
\hat{t}}_{k}\sin\alpha \\
{\bf \hat{a}}_{k} & = & {\bf \hat{t}}_{k} \cos\alpha + {\bf
\hat{r}}_{k}\sin\alpha  \end{array} \; \; .
\eeq
We remark that
\beq
{\bf \hat{a}}_{k} = \frac{{\bf \vec{r}}_{k+1} - {\bf \vec{r}}_{k}}{|{\bf
\vec{r}}_{k+1} - {\bf \vec{r}}_{k} |} \; \; .
\eeq
We define the normal vector ${\bf \hat{n}}_{k}$ by ${\bf \hat{n}}_{k}={\bf
\hat{r}}_{k}\times {\bf \hat{t}}_{k}$, where the cross-product $\times$ is
consistent with the given orientation of the ambient ${\bf R}^{3}$. The three
vectors $({\bf \hat{r}}_{k},{\bf \hat{t}}_{k},{\bf \hat{n}}_{k})$
form the orthonormal basis of ${\bf R}^{3}$ in the point ${\bf
\vec{r}}_{k}$ of the discrete curve.

The oriented tangent plane $\tau_{k}$ is given by the ordered pair of
vectors $({\bf \hat{t}}_{k},{\bf \hat{n}}_{k})$. Since ${\bf \hat{r}}_{k} \bot
\tau_{k}$ and ${\bf \hat{r}}_{k} \bot {\bf \hat{t}}'_{k-1}$, then ${\bf
\hat{t}}'_{k-1} \in \tau_{k}$. By $\varphi_{k-1}$ ($|\varphi_{k-1}|< \pi $)
we denote the angle between
${\bf \hat{t}}'_{k-1}$ and ${\bf \hat{t}}_{k}$ measured according to the
orientation of the plane $\tau_{k}$; $\varphi_{k-1}$ is the angle of geodesic
curvature. We remark that $\varphi_{k-1}$ is also the angle between the
planes $\pi_{k-1}$ and $\pi_{k}$.

\bigskip

%\epsffile{02.ps}

\bigskip
\bigskip

\noindent In the continuous limit: $\Delta \rightarrow 0$ and
$\Delta k \rightarrow s$, where $s$ is the arc-length parameter of
the corresponding
curve, the angle of geodesic curvature reduces to the familiar geodesic
curvature $\kappa(s)$ in the following way:
\beq
\frac{\vf _{k}}{\Delta} \stackrel{\Delta \rightarrow 0}{\longrightarrow} \kappa
(s) \; \; .
\eeq
\noindent The transition from the basis $({\bf \hat{r}}_{k},{\bf \hat{t}}_{k},
{\bf \hat{n}}_{k})$ to the basis $({\bf \hat{r}}_{k+1},{\bf \hat{t}}_{k+1},{\bf
\hat{n}}_{k+1})$ can be obtained by the superposition of two rotations:

\medskip

\noindent i) rotation in the plane $\pi_{k}$ of the angle $\lambda\Delta =
2\alpha$
\[({\bf \hat{r}}_{k},{\bf \hat{t}}_{k},{\bf \hat{n}}_{k}) \longrightarrow
({\bf \hat{r}}_{k+1},{\bf \hat{t}}'_{k},{\bf \hat{n}}_{k})
\]
ii) rotation in the plane $\tau_{k+1}$ of the angle $\vf_{k}$
\[({\bf \hat{r}}_{k+1},{\bf \hat{t}}'_{k},{\bf \hat{n}}_{k}) \longrightarrow
({\bf \hat{r}}_{k+1},{\bf \hat{t}}_{k+1},{\bf \hat{n}}_{k+1}) \; \; .
\]
The full rotation can be written in matrix form as:
\beq \label{eq:mlp}
\left[ \begin{array}{c}
{\bf \hat{r}}_{k+1} \\ {\bf \hat{t}}_{k+1} \\ {\bf \hat{n}}_{k+1}
 \end{array} \right] = \left[ \begin{array}{ccc}
\cos (\lambda\Delta)  &  \sin (\lambda\Delta)  & 0    \\
-\cos\varphi_{k} \sin(\lambda\Delta) & \cos\varphi_{k} \cos(\lambda\Delta) &
\sin\varphi_{k}\\
 \sin\varphi_{k}\sin(\lambda\Delta) & - \sin\varphi_{k} \cos(\lambda\Delta) &
\cos\varphi_{k}
\end{array} \right] \left[ \begin{array}{c}
{\bf \hat{r}}_{k} \\ {\bf \hat{t}}_{k} \\ {\bf \hat{n}}_{k}
\end{array} \right] \; \; .
\eeq

\subsection{The spinor representation of the linear problem}

To obtain the $2\times 2$ version of the linear problem (\ref{eq:mlp}) we use
the standard representation of rotations in ${\bf R}^{3}$ in terms of
$SU(2)$-valued matrices \cite{LaMi}. From this point of view ${\bf R}^{3}$ can
be considered as the space spanned by the matrices $\ep_{1} = i \sigma_{2} $,
$\ep_{2} = -i\sigma_{1} $, $\ep_{3} = i\sigma_{3}$ (here $\sigma_{l}$ denote
the standard Pauli matrices):
\beq
{\bf \vec{v}}\in {\bf R}^{3} \; \Leftrightarrow \; \sigma({\bf \vec{v}}) =
\sum_{l=1}^{3}{\rm v}^{l}\ep_{l}  \; \; ; \; \; || {\bf \vec{v}} ||^{2} = \det
\sigma({\bf \vec{v}}) \; \; ,
\eeq
where ${\rm v}^{l}$ are the Cartesian coordinates of the vector ${\bf
\vec{v}}$.
We recall that any rotation $O$ in ${\bf R}^{3}$ can be represented by a matrix
$S\in SU(2)$ by the formula
\beq
{\bf \vec{w}}= O{\bf \vec{v}} \Leftrightarrow \sigma({\bf \vec{w}}) = S^{-1}
\sigma({\bf \vec{v}}) S \; \; .
\eeq
Suppose we are given a sequence of orthonormal frames ${\cal F}_{k}=({\bf
e}_{k}, {\bf f}_{k}, {\bf g}_{k})$, $k\in {\bf Z}$, and the
corresponding matrices $S_{k}$
\beq
{\cal F}_{k} = S_{k}^{-1} {\cal E} S_{k} \; \; ,
\eeq
where ${\cal E} = (\ep_{1}, \ep_{2}, \ep_{3})$.
When matrices $A_{k}$ are defined by the relation
\beq
S_{k+1}=A_{k} S_{k}
\eeq
then they satisfy also
\beq
S_{k}{\cal F}_{k+1}S_{k}^{-1} = A_{k}^{-1}{\cal E}A_{k} \; \; .
\eeq
For example, the matrix
\beq
A_{k} = \left[ \begin{array}{cc}
{\rm e}^{i \alpha} & 0 \\
0 & {\rm e}^{-i\alpha} \end{array} \right]
\eeq
represents the rotation in the plane $({\bf e}_{k}, {\bf f}_{k})$ of the angle
$2\alpha$, and the matrix
\beq
A_{k} = \left[ \begin{array}{cc}
\cos(\vf/2) & \sin(\vf/2) \\
-\sin(\vf/2) & \cos(\vf/2) \end{array} \right]
\eeq
represents the rotation in the plane $({\bf f}_{k}, {\bf g}_{k})$ of the angle
$\vf$.

When the sequence of orthonormal frames is the sequence of the Frenet frames in
the points of the discrete curve: ${\cal F}_{k}=({\bf \hat{r}}_{k},{\bf
\hat{t}}_{k},{\bf \hat{n}}_{k})$, then the corresponding spin matrices $S_{k}$
are subjected to the relation
\begin{eqnarray} \label{eq:2sp}
S_{k+1} & = & \left[ \begin{array}{cc}
{\rm e}^{i\alpha}\cos(\vf_{k}/2)  & {\rm e}^{- i\alpha} \sin(\vf_{k}/2) \\
-{\rm e}^{i\alpha} \sin(\vf_{k}/2) & {\rm e}^{-i\alpha}\cos(\vf_{k}/2)
\end{array} \right] S_{k} \nonumber \\
 & = & \frac{1}{\sqrt{1+q_{k}^{2}}} \left[ \begin{array}{cc}
\zeta  &\zeta^{-1} q_{k}   \\
-\zeta q_{k} & \zeta^{-1} \end{array} \right] S_{k} =: A_{k} S_{k} \; \; ,
\end{eqnarray}
with
\beq
q_{k}=\tan (\vf_{k}/2) \; \; , \; \; \zeta = {\rm e}^{i\alpha} \; \; .
\eeq
We remark that this spectral problem is equivalent to the AL spectral problem
\beq
\tilde{S}_{k+1} = \left[ \begin{array}{cc}
\zeta  & q_{k} \\
-q_{k} & \zeta^{-1} \end{array} \right] \tilde{S}_{k} \; \;
\eeq
in the reduction $r_{k} = - q_{k} \in {\bf R}$.
To show this we relax the (unessential) condition $\det S_{k} =1$ to remove the
$\cos (\vf_{k}/2)$ diagonal term and apply a gauge transformation which
corresponds to the change of the basis of the plane $\pi_{k}$ from  $({\bf
\hat{r}}_{k}, {\bf \hat{t}}_{k})$ to  $({\bf \hat{r}}'_{k}, {\bf
\hat{a}}_{k})$: ${\cal F}_{k}=({\bf \hat{r}}_{k},{\bf \hat{t}}_{k},{\bf
\hat{n}}_{k}) \rightarrow \tilde{{\cal F}}_{k}=({\bf \hat{r}}'_{k},{\bf
\hat{a}}_{k}, {\bf \hat{n}}_{k})$. Then
\beq
\tilde{{\cal F}}_{k} = \tilde{S}_{k}^{-1} {\cal E} \tilde{S}_{k} \; \; ,
\eeq
where
\beq
\tilde{S}_{k}= c_{k}\left[ \begin{array}{cc} {\rm e}^{i\alpha/2}  & 0 \\
0 & {\rm e}^{-i\alpha/2} \end{array} \right] S_{k}\left[ \begin{array}{cc}
{\rm e}^{-i\alpha/2}  & 0 \\
0 & {\rm e}^{i\alpha/2} \end{array} \right] \; \; ,
\eeq
and
\beq
c_{k+1}=\frac{c_{k}}{\cos(\vf_{k}/2)} = c_{k}\sqrt{1+q_{k}^{2}} \; \; .
\eeq

\subsection{The integrable motions on $S^{2}(R)$}
\label{sec:kin2}

Since the motion takes place on the sphere, the velocity field is tangent to
it:
\beq
{\bf \vec{r}}_{k,t} = \tilde{V}_{k}{\bf \hat{t}}_{k} +\tilde{U}_{k}{\bf
\hat{n}}_{k} \; \; .
\eeq
Futhermore the evolution of the Frenet frame is described by an antisymmetric
matrix:
\beq \label{eq:evF}
\left[ \begin{array}{c}
{\bf \hat{r}}_{k} \\ {\bf \hat{t}}_{k} \\ {\bf \hat{n}}_{k}
\end{array} \right] _{,t} = \left[ \begin{array}{ccc}
0       & \lambda \tilde{V}_{k} & \lambda \tilde{U}_{k} \\
-\lambda \tilde{V}_{k} &   0    & A_{k}  \\
-\lambda \tilde{U}_{k} & -A_{k}  &   0
\end{array} \right]
\left[ \begin{array}{c}
{\bf \hat{r}}_{k} \\ {\bf \hat{t}}_{k} \\ {\bf \hat{n}}_{k}
\end{array} \right] \; \; ,
\eeq
or equivalently, by the evolution of the matrix $S_{k}$:
\beq \label{eq:ev2}
S_{k,t} = \frac{1}{2}\left[ \begin{array}{cc}
i\lambda \tilde{V}_{k} & A_{k} + i \lambda \tilde{U}_{k} \\
- A_{k} + i \lambda \tilde{U}_{k} & i\lambda \tilde{V}_{k}
\end{array} \right] S_{k} =: T_{k}S_{k} \; \; .
\eeq
Using {\it Property 2}' (i.e. $\Delta_{,t} = 0$), the compatibility condition
\beq
E\partial_{t} = \partial_{t} E \; \; ,
\eeq
where $E$ is the shift operator along the chain:
\beq
E f_{k} = f_{k+1} \; \; ,
\eeq
applied to the linear problems (\ref{eq:mlp}) and (\ref{eq:evF}), or
(\ref{eq:2sp}) and (\ref{eq:ev2}), specifies $A_{k}$ in terms of the velocity:
\begin{eqnarray}
A_{k} & = &  \frac{\lambda}{\sin(\lambda\Delta)}\left[
\frac{2q_{k}}{1+q^{2}_{k}}
\tilde{V}_{k+1} + \frac{1-q^{2}_{k}}{1+q^{2}_{k}} \tilde{U}_{k+1} -
\cos(\lambda\Delta) \tilde{U}_{k} \right] =  \\
 & = &  \frac{\lambda}{\sin(\lambda\Delta)} \left[ \sin \vf _{k}\tilde{V}_{k+1}
+ \cos \vf _{k}\tilde{U}_{k+1} -
\cos(\lambda\Delta) \tilde{U}_{k} \right] \nonumber
\end{eqnarray}
and yields the kinematics
\beq \label{eq:kin1}
2q_{k,t} = -2\cos(\lambda\Delta)\left( U_{k+1} + q_{k}V_{k+1} \right) + (1+
q_{k}^{2}) \left[ U_{k} + U_{k+2} + q_{k+1} \left( V_{k+1} + V_{k+2} \right)
\right] \; \; ,
\eeq
\beq \label{eq:kin2}
(1-q_{k}^{2}) V_{k+1} - (1+q_{k}^{2}) V_{k} = 2q_{k} U_{k+1} \; \; \; \;
( \cos \vf_{k}V_{k+1} -  V_{k}= \sin \vf_{k}  U_{k+1} )\;  ,
\eeq
where
\beq
V_{k} = \frac{\lambda}{\sin(\lambda\Delta)}\tilde{V}_{k} \; \; , \; \;
U_{k} = \frac{\lambda}{\sin(\lambda\Delta)}\tilde{U}_{k} \; \; .
\eeq
Using (\ref{eq:kin2}), equation (\ref{eq:kin1}) can be rewritten in the
following convenient form
\beq \label{eq:kin-rec}
2q_{k,t} = -2\cos(\lambda\Delta)\left( U_{k+1} + q_{k}V_{k+1} \right) +
{\cal R} \left( U_{k+1} +q_{k} V_{k+1} \right)
\eeq
in terms of the recursion operator
\beq
{\cal R} := (1+q_{k}^{2})\left[ E + E^{-1} + 2( q_{k+1} E - q_{k-1}) (E-1)^{-1}
\frac{q_{k}}{1+q^{2}_{k}} \right]
\eeq
which is a real reduction of the recursion operator obtained in \cite{ChL},
associated with the AL spectral problem (\ref{eq:dsp}).
The operator $(E-1)^{-1}$, formal inverse of $(E-1)$, takes the following
explicit form:
\beq
(E-1)^{-1} f_{k} = \sum_{j=k_{0}}^{k-1} f_{j} \: + b \; \; ,
\eeq
where the choice of the constants $k_{0}$ and $b$ depends on the space of
functions $f_{k}$ under consideration.

Equations (\ref{eq:kin-rec})(\ref{eq:kin2}) show that the kinematics of the
curve
depends explicitly on the radius of the sphere through $\cos(\lambda\Delta)$;
a drastic way to satisfy {\it
Property 3}' is to choose the combination $(U_{k+1} + q_{k} V_{k+1})$ as the
real eigenfunction of the recursion operator ${\cal R}$ for the eigenvalue
$2\cos(\lambda\Delta)$, i.e.:
\beq
U_{k+1} + q_{k} V_{k+1} = \Psi_{k}(\lambda) \; \; ,
\eeq
where
\beq
{\cal R}\Psi_{k}(\lambda) = 2\cos(\lambda\Delta) \Psi_{k}(\lambda) \ \; .
\eeq
This velocity field, which generates the rigid motion $q_{k,t}=0$ of the
discrete curve on the sphere of radius $1/\lambda$, can be used to generate
more complicated dynamics. For instance, if, on the sphere of radius
$1/\lambda$, one uses
the velocity field which would give rise to the rigid motion on the sphere of
radius $1/\lambda_{0} \not= 1/\lambda$, normalized in the following way
\beq
U_{k+1} + q_{k} V_{k+1} = \frac{\Psi_{k}(\lambda_{0})}{\cos(\lambda_{0}\Delta)
 - \cos(\lambda\Delta)} \; \; ,
\eeq
then one would satisfy {\it Property 3}', obtaining an example of integrable
discrete dynamics with sources (with a singular dispersion relation):
\beq
q_{k,t} = \Psi_{k}(\lambda_{0}) \; \; \Leftrightarrow \; \; [{\cal R} -
2\cos(\lambda_{0}\Delta)] q_{k,t} = 0 \; \; .
\eeq
If, in particular, $\lambda_{0} = 0$, then
\beq
{\bf \vec{v}}_{k} = \frac{\sin(\lambda\Delta)}{\lambda[1-\cos(\lambda\Delta)]}
(V_{k} {\bf \hat{t}}_{k} + U_{k} {\bf \hat{n}}_{k} ) \; \; ,
\eeq
where
\begin{eqnarray}
\label{eq:vel_sG}
U_{k+1} - U_{k} + q_{k} ( V_{k+1} + V_{k} ) & = & 0  \; \; , \nonumber \\
V_{k+1} - V_{k} - q_{k} ( U_{k+1} + U_{k} ) & = & 0  \; \; ,
\end{eqnarray}
and
\beq
\label{eq:d_sG}
q_{k,t}=\Psi_{k}(0) = \frac{1}{2} (1+ q_{k}^{2}) ( U_{k+1} + U_{k} ) \;  \;
\Leftrightarrow \; \; \vf_{k,t} = U_{k+1} + U_{k} \; \; .
\eeq
Equations (\ref{eq:vel_sG}), satisfied by the velocity field which gives rise
to the rigid motion on the plane, can be integrated to
\beq
U_{k} = c \sin\Omega_{k} \; \; , \; \; V_{k} = - c \cos\Omega_{k} \; \; ,
\eeq
where $\Omega_{k}$ is the (incomplete) total curvature:
\beq
\Omega_{k} = \sum_{l}^{k-1}{\vf_{l}} \; \; , \; \; ( \Omega_{k+1} - \Omega_{k}
 = \vf_{k} ) \; .
\eeq
Consequently equation (\ref{eq:d_sG}) becomes the well-known discrete
sine-Gordon (SG) equation \cite{Orf}\cite{Le}:
\beq
\label{eq:d_sG_2}
\vf_{k,t} = c (\sin\Omega_{k+1} + \sin\Omega_{k} ) \; \; \Leftrightarrow
( \Omega_{k+1} - \Omega_{k} )_{,t} = c (\sin\Omega_{k+1} + \sin\Omega_{k}) \; ,
\eeq
which can also be written in the following form
\beq
\label{eq:nd_sG}
(u_{k+1}  - u_{k})_{,t} = c \sin(u_{k+1}  + u_{k})
\eeq
through the change of variables
\beq
u_{k+1} + u_{k} = \Omega_{k} \; \; , \; \;
(u_{k} =\sum_{j}^{1}{\vf_{k-2j}}) \; .
\eeq
Using the well-known connection between discrete dynamics and B\"{a}cklund
transformations (BT) \cite{LeBen}, we observe that equation (\ref{eq:nd_sG})
can be viewed as the celebrated BT of the SG equation $u_{,tx} = \sin u$
\cite{Bac}, interpreting $u_{k}$ and $u_{k+1}$ as two different solutions of
it. Therefore equation (\ref{eq:d_sG_2}) has two different geometric meanings:
i) it describes the integrable dynamics of a sequence of points on the sphere,
subjected to a velocity field which would give rise to a rigid motion on the
plane, and ii) it provides an elegant way to construct a new (pseudo)spherical
surface from a given one.

\medskip

The linear dependence on $\cos(\lambda\Delta)$ in equation (\ref{eq:kin-rec})
suggests that {\it Property 3}' can be satisfied also looking for velocity
fields given by suitable Laurent expansions in powers of $\cos(\lambda\Delta)$:
\beq \label{eq:expan}
\left[ \begin{array}{c} V_{k} \\ U_{k} \end{array} \right] =
\sum_{j=-m_{1}}^{m_{2}} (\cos(\lambda\Delta))^{j} \left[ \begin{array}{c}
V_{k}^{(m-j)} \\ U_{k}^{(m-j)} \end{array} \right]  \; \; , \; m_{1} , m_{2}
\geq 0 \; \; .
\eeq
Substituting equation (\ref{eq:expan}) into (\ref{eq:kin-rec}) and requiring
independence of $\cos(\lambda\Delta)$, one obtains a class of integrable
dynamics in the form
\beq
g({\cal R}) q_{k,t} = h({\cal R}) (1+q_{k}^{2}) (q_{k+1} - q_{k-1}) \; \; ,
\eeq
where $g$ and $h$ are arbitrary entire functions of their arguments. The
simplest examples are the following:

\medskip

\noindent i) If $g=h=1$ we have
\beq
{\bf \vec{v}}_{k} = \frac{\sin(\lambda\Delta)}{\lambda}\left[ {\bf \hat{t}}_{k}
- q_{k-1} {\bf \hat{n}}_{k} \right] \; \; ,
\eeq
\beq
q_{k,t} = (1+q_{k}^{2}) (q_{k+1} - q_{k-1}) := K_{k}^{(1)} \; \;
\Leftrightarrow \vf_{k,t} = \tan(\frac{\vf_{k+1}}{2})-
\tan(\frac{\vf_{k-1}}{2})
\; \; .
\eeq
In this case the velocity field ${\bf \vec{v}}_{k}$ admits the simple geometric
construction shown in the figure below, in the case of the plane ($\lambda
\rightarrow 0$).

%\epsffile{03.ps}

\bigskip
\bigskip

\noindent ii) If $g=1$ and $h(x)=x$ we have
\beq
{\bf \vec{v}}_{k} = \frac{4 \sin(\lambda\Delta)}{\lambda}\left\{ \left(
\cos(\lambda\Delta) + q_{k} q_{k-1} \right) {\bf \hat{t}}_{k} +
\left( -q_{k-1}\cos(\lambda\Delta) + \frac{1}{2}(1+q_{k-1}^{2}) (q_{k} -
q_{k-2}) -  q_{k-1}^{2}q_{k} \right) {\bf \hat{n}}_{k} \right\} \; ,
\eeq
\beq
q_{k,t} = (1+q_{k}^{2}) \left[ (1+q_{k+1}^{2})q_{k+2} -(1+ q_{k-1}^{2})q_{k-2}
+ q_{k}(q_{k+1}^{2} - q_{k-1}^{2}) \right] := K_{k}^{(2)} \; \; .
\eeq
As it was observed in \cite{AL}, the linear combination
\beq
q_{k,t} = K_{k}^{(2)}  - 2 K_{k}^{(1)}
\eeq
of the first two flows reduces, in the continuous limit, to the mKdV equation
\beq
\kappa_{,t'} = \kappa_{,sss} + \frac{3}{2} \kappa^{2} \kappa_{,s} \; \; ,
\; \;  t' = 2 \Delta^{3} t  \; \; .
\eeq
\noindent iii) If $g(x)=x$ and $h=0$, then
\beq
{\bf \vec{v}}_{k} = \frac{\tan(\lambda\Delta)}{\lambda}\left[ V_{k}^{(-1)}{\bf
\hat{t}}_{k} + U_{k}^{(-1)} {\bf \hat{n}}_{k}\right] \; \; ,
\eeq
\beq
q_{k,t} = -\frac{1}{2}(1+q_{k}^{2})(U_{k+1}^{(-1)} - U_{k-1}^{(-1)}) \; \; ,
\eeq
where the normal velocity field satisfies the difference equation
\beq
\left( q_{k+1} + \frac{1}{q_{k+1}} \right) U_{k+2}^{(-1)} +
\left( q_{k} - \frac{1}{q_{k}} \right) U_{k+1}^{(-1)} -
\left( q_{k+1} - \frac{1}{q_{k+1}} \right) U_{k}^{(-1)} -
\left( q_{k} + \frac{1}{q_{k}} \right) U_{k-1}^{(-1)} = 0 \; \; ,
\eeq
and the tangent velocity
\beq
2V_{k}^{(-1)} = - \left( q_{k} + \frac{1}{q_{k}} \right) U_{k+1}^{(-1)} +
\left( q_{k} - \frac{1}{q_{k}} \right) U_{k-1}^{(-1)}  \; \; .
\eeq
We remark that this integrable dynamics corresponds to the velocity field
\beq
U_{k+1} + q_{k} V_{k+1} = -\frac{\Psi_{k}(\lambda_{0})}{\cos(\lambda\Delta)}
\; \; ,
\eeq
which would give rise to a rigid motion on the sphere of radius $1/\lambda_{0}
= 2\Delta/\pi$.

\section{The motion of a discrete curve in $S^{3}(R)$}

\subsection{The discrete curve in $S^{3}(R)$}
The plane $\pi_{k}$ is, as in the previous Section, the oriented plane given by
the ordered pair of vectors $({\bf \vec{r}}_{k} \, , \, {\bf \vec{r}}_{k+1})$.
Similarly, the unit vectors ${\bf \hat{t}}_{k}$ and ${\bf \hat{t}}'_{k}$ are
tangent to the big circle passing through the points ${\bf \vec{r}}_{k}$ and
${\bf \vec{r}}_{k+1}$.

The tangent plane $\tau_{k}$ in the point ${\bf \vec{r}}_{k}$ to the discrete
curve
is defined by the two vectors ${\bf \hat{t}}'_{k-1}$ and ${\bf \hat{t}}_{k}$.
The angle of geodesic curvature $\vf_{k}$ is the angle between the vectors
${\bf \hat{t}}'_{k}$ and ${\bf \hat{t}}_{k+1}$. The orientation of the plane
$\tau_{k+1}$ is choosen by the condition $0<\vf_{k} < \pi$. The normal vector
${\bf \hat{n}}_{k}$ is the unit vector of $\tau_{k}$ orthogonal to ${\bf
\hat{t}}_{k}$ and such that the pair $({\bf \hat{t}}_{k},{\bf \hat{n}}_{k})$
gives correct orientation of the plane $\tau_{k}$.

The binormal vector ${\bf \hat{b}}_{k}={\bf \hat{r}}_{k}\times {\bf
\hat{t}}_{k}\times{\bf \hat{n}}_{k}$, together with the normal vector ${\bf
\hat{n}}_{k}$, define the normal plane $\nu_{k}$ in point ${\bf
\vec{r}}_{k}$. The orientation of the plane $\nu_{k}$ is given by the ordering
$({\bf \hat{n}}_{k},{\bf \hat{b}}_{k})$.
We first remark that the vectors ${\bf \hat{r}}_{k+1}$ and ${\bf \hat{t}}'_{k}$
span the plane $\pi_{k}$; moreover, since ${\bf \hat{t}}'_{k}\in \tau_{k+1}$,
it follows that ${\bf \hat{b}}_{k+1}\bot {\bf \hat{t}}'_{k}$ which, together
with the obvious condition ${\bf \hat{b}}_{k+1}\bot {\bf \hat{r}}_{k+1}$,
imply that ${\bf \hat{b}}_{k+1}\in \nu_{k}=\pi_{k}^{\bot}$. The angle
$\theta_{k}$ between the vectors ${\bf \hat{b}}_{k}$ and ${\bf \hat{b}}_{k+1}$
(which belong to the same plane $\nu_{k}$), measured according to the
orientation of the plane $\nu_{k}$, is called the angle of (the geodesic)
torsion of the curve in point ${\bf \vec{r}}_{k}$. In the case of ${\bf R}^{3}$
it is ilustrated on the figure below

%\epsffile{04.ps}

\bigskip
\bigskip

\noindent In the continuous limit, the angle of torsion $\theta_{k}$ reduces
to the
familiar torsion $\tau(s)$ of the corresponding continuous curve in the
following way:
\beq
\frac{\theta _{k}}{\Delta} \stackrel{\Delta \rightarrow 0}{\longrightarrow}
\tau(s) \; \; .
\eeq

\subsection{The Hasimoto transformation and the spectral problem}

As in the case of the continuous curve, it is convenient to change the basis of
the normal plane $\nu_{k}$ through a rotation of the angle $\sigma_{k-1} =
(E-1)^{-1}\theta_{k} = \sum_{l}^{k-1}\theta_{l}$.
\beq
{\bf \hat{n}}_{k}^{1} = \cos \sigma_{k-1} \: {\bf \hat{n}}_{k} - \sin
\sigma_{k-1}  \: {\bf \hat{b}}_{k} \; \; ,
\eeq
\[
{\bf \hat{n}}_{k}^{i} = \sin \sigma_{k-1} \: {\bf \hat{n}}_{k} + \cos
\sigma_{k-1}  \: {\bf \hat{b}}_{k} \; \; .
\]
This change of basis corresponds to a partial "integration" of the Frenet
equations in the normal plane, since the vectors ${\bf \hat{n}}_{k}^{1}$, ${\bf
\hat{n}}_{k}^{i}$ do not vary from the point of view of the normal plane (this
is the discrete analog of the parallel transport in the normal bundle).

As in the case of a continuous curve, it is convenient to interprete any vector
of the normal plane $\nu_{k}$ as the complex number
\beq
{\bf \vec{\phi}}_{k} = {\rm Re} \phi_{k} \: {\bf \hat{n}}_{k}^{1} + {\rm Im}
\phi_{k} \:
{\bf \hat{n}}_{k}^{i} \; \; \Leftrightarrow \; \; \phi_{k} = {\rm Re} \phi_{k}
+ i\:
{\rm Im} \phi_{k} \; \; .
\eeq
The transition from the basis $({\bf \hat{r}}_{k},{\bf \hat{t}}_{k},{\bf
\hat{n}}_{k}^{1},{\bf \hat{n}}_{k}^{i})$ to the basis $({\bf
\hat{r}}_{k+1},{\bf \hat{t}}_{k+1},{\bf \hat{n}}_{k+1}^{1},{\bf
\hat{n}}_{k+1}^{i} )$ can be obtained by the superposition of the following
five rotations:

\medskip

\noindent i) rotation in the plane $\nu_{k}$ of the angle $\sigma_{k-1}$
\[({\bf \hat{r}}_{k},{\bf \hat{t}}_{k},{\bf \hat{n}}_{k}^{1},{\bf
\hat{n}}_{k}^{i}) \longrightarrow ({\bf \hat{r}}_{k},{\bf \hat{t}}_{k},{\bf
\hat{n}}_{k},{\bf \hat{b}}_{k})
\]
ii) rotation in the plane $\pi_{k}$ of the angle 2$\alpha=\lambda\Delta$
\[({\bf \hat{r}}_{k},{\bf \hat{t}}_{k},{\bf \hat{n}}_{k},{\bf \hat{b}}_{k})
\longrightarrow ({\bf \hat{r}}_{k+1},{\bf \hat{t}}'_{k},{\bf \hat{n}}_{k},{\bf
\hat{b}}_{k})
\]
iii) rotation in the plane $\nu_{k}$ of the angle $\theta_{k}$
\[({\bf \hat{r}}_{k+1},{\bf \hat{t}}_{k}',{\bf \hat{n}}_{k},{\bf \hat{b}}_{k})
\longrightarrow ({\bf \hat{r}}_{k+1},{\bf \hat{t}}'_{k},{\bf \hat{n}}_{k}',{\bf
\hat{b}}_{k+1})
\]
iv) rotation in the plane $\tau_{k+1}$ of the angle $\vf_{k}$
\[({\bf \hat{r}}_{k+1},{\bf \hat{t}}'_{k},{\bf \hat{n}}_{k}',{\bf
\hat{b}}_{k+1}) \longrightarrow
({\bf \hat{r}}_{k+1},{\bf \hat{t}}_{k+1},{\bf \hat{n}}_{k+1},{\bf
\hat{b}}_{k+1})
\]
v) rotation in the plane $\nu_{k+1}$ of the angle $-\sigma_{k}$
\[({\bf \hat{r}}_{k+1},{\bf \hat{t}}_{k+1},{\bf \hat{n}}_{k+1},{\bf
\hat{b}}_{k+1}) \longrightarrow ({\bf \hat{r}}_{k+1},{\bf \hat{t}}_{k+1},{\bf
\hat{n}}_{k+1}^{1},{\bf \hat{n}}_{k+1}^{i}) \; .
\]
Rotations in ${\bf R}^{4}$ can be represented in a similar form as in
the previous Section. A fixed Cartesian basis in ${\bf R}^{4}$ can be
identified with the following $2\times 2$ matrices ${\cal E}= (\ep_{0}={\rm
Id}, \ep_{1} = i\sigma_{3}, \ep_{2}=i\sigma_{1}, \ep_{3} = -i\sigma_{2} )$.
\beq
{\bf \vec{v}}\in {\bf R}^{4} \; \Leftrightarrow \; \sigma({\bf \vec{v}}) =
\sum_{l=0}^{3}{\rm v}^{l}\ep_{l}  \; \; ; \; \; || {\bf \vec{v}} ||^{2} = \det
\sigma({\bf \vec{v}}) \; \; .
\eeq
Any rotation $O$ can be represented in terms of two matrices $P$ and $S$ of
the $SU(2)$ group in the following way
\beq
{\bf \vec{w}}= O{\bf \vec{v}} \Leftrightarrow \sigma({\bf \vec{w}}) = P^{-1}
\sigma({\bf \vec{v}}) S \; \; .
\eeq
When we are given a sequence of orthonormal frames ${\cal F}_{k}=({\bf
e}_{k}, {\bf f}_{k}, {\bf g}_{k}, {\bf h}_{k})$ and the
corresponding matrices $P_{k}, S_{k}$
\beq
{\cal F}_{k} = P_{k}^{-1} {\cal E} S_{k} \; \;
\eeq
then we define matrices $A_{k}$ and $B_{k}$ by the relations
$S_{k+1}=A_{k} S_{k}$ and $P_{k+1}=B_{k} P_{k}$; moreover
\beq
P_{k}{\cal F}_{k+1}S_{k}^{-1} = B_{k}^{-1}{\cal E}A_{k} \; \; .
\eeq
If the rotation takes place in the space perpendicular to the first vector of
the basis ${\cal F}_{k}$ (i.e. ${\bf e}_{k+1}={\bf e}_{k}$) then $A_{k}=B_{k}$.
For example the matrices
\beq
A_{k} = B_{k} = \left[ \begin{array}{cc}
{\rm e}^{-i\theta/2} & 0 \\
0 & {\rm e}^{i\theta/2} \end{array} \right]
\eeq
represent the rotation in the plane $({\bf g}_{k}, {\bf h}_{k})$ of the angle
$\theta$, and the matrices
\beq
A_{k} = B_{k} =\left[ \begin{array}{cc}
\cos(\vf/2) & \sin(\vf/2) \\
-\sin(\vf/2) & \cos(\vf/2) \end{array} \right]
\eeq
represent the rotation in the plane $({\bf f}_{k}, {\bf g}_{k})$ of the angle
$\vf$.

We will also need the representation of the rotation in the plane $({\bf
e}_{k}, {\bf f}_{k})$ of the angle $2\alpha$,  given by matrices
\beq
A_{k} = \left[ \begin{array}{cc}
{\rm e}^{i\alpha} & 0 \\
0 & {\rm e}^{-i\alpha} \end{array} \right] \; \; , \; \;
B_{k} = \left[ \begin{array}{cc}
{\rm e}^{-i\alpha} & 0 \\
0 & {\rm e}^{i\alpha} \end{array} \right] \; \; .
\eeq
We are now in the position to calculate the transition matrices when the
sequence
of the orthonormal frames is the sequence of the frames along the
discrete curve ${\cal F}_{k}=({\bf \hat{r}}_{k},{\bf \hat{t}}_{k},{\bf
\hat{n}}_{k}^{1},{\bf \hat{n}}_{k}^{i})$. The corresponding spin matrices
$P_{k}, S_{k}$ are subjected to the relations
\begin{eqnarray}
P_{k+1} & = & \left[ \begin{array}{cc}
{\rm e}^{-i\alpha}\cos(\vf_{k}/2)  & {\rm e}^{i\alpha}\sin(\vf_{k}/2) {\rm
e}^{i\sigma_{k}} \\
-{\rm e}^{-i\alpha}\sin(\vf_{k}/2){\rm e}^{-i\sigma_{k}} & {\rm
e}^{i\alpha}\cos(\vf_{k}/2)
\end{array} \right] P_{k} \nonumber \\
 & = & \frac{1}{\sqrt{1+|q_{k}|^{2}}} \left[ \begin{array}{cc}
\zeta^{-1}  &\zeta q_{k}   \\
-\zeta^{-1}\bar{q}_{k} & \zeta \end{array} \right] P_{k} =: B_{k} P_{k} \; \; ,
\end{eqnarray}

\begin{eqnarray} \label{eq:S}
S_{k+1} & = & \left[ \begin{array}{cc}
{\rm e}^{i\alpha}\cos(\vf_{k}/2)  & {\rm e}^{-i\alpha}\sin(\vf_{k}/2) {\rm
e}^{i\sigma_{k}} \\
-{\rm e}^{i\alpha}\sin(\vf_{k}/2){\rm e}^{-i\sigma_{k}} & {\rm
e}^{-i\alpha}\cos(\vf_{k}/2)
\end{array} \right] S_{k} \nonumber \\
 & = & \frac{1}{\sqrt{1+|q_{k}|^{2}}} \left[ \begin{array}{cc}
\zeta  &\zeta^{-1} q_{k}   \\
-\zeta\bar{q}_{k} & \zeta^{-1} \end{array} \right] S_{k} =: A_{k} S_{k} \; \; ,
\end{eqnarray}
where
\beq
\label{eq:q}
q_{k}=\tan (\vf_{k}/2){\rm e}^{i\sigma_{k}} \; \; , \; \; \zeta = {\rm
e}^{i\alpha} \; \; .
\eeq
We remark that the transformation (\ref{eq:q}) between the curvature and
torsion angles and the complex field $q_{k}$ is the discrete analogue of the
Hasimoto transformation, introduced for continuous curves in \cite{Hasimoto}.
We also remark that $B_{k} = A_{k} {\rm diag}[\zeta^{-2}, \zeta^{2}]$.

As before the linear problem can be transformed into the AL spectral problem:
\beq
\tilde{S}_{k+1} = \left[ \begin{array}{cc}
\zeta  & q_{k} \\
-\bar{q}_{k} & \zeta^{-1} \end{array} \right] \tilde{S}_{k}
\; \; , \; \;
\tilde{P}_{k+1} = \left[ \begin{array}{cc}
\zeta^{-1}  & q_{k} \\
-\bar{q}_{k} & \zeta \end{array} \right] \tilde{P}_{k} \; \; ,
\eeq
(but now in the reduction $r_{k} = -\bar{q}_{k}$) relaxing the conditions
$\det S_{k} =\det P_{k} =1$ (but keeping still $\det S_{k} = \det P_{k}$)
and changing the basis $({\bf \hat{r}}_{k}, {\bf \hat{t}}_{k})$ of the plane
$\pi_{k}$ to the new basis $({\bf \hat{r}}'_{k}, {\bf \hat{a}}_{k})$.

\subsection{The integrable motions in $S^{3}(R)$}

Proceeding as in Section \ref{sec:kin2} we have that
\beq \label{eq:kin_r} {\bf \vec{r}}_{k,t} =
\frac{\sin(\lambda\Delta)}{\lambda} \left( V_{k} {\bf \hat{t}}_{k} + {\bf
\vec{\phi}}_{k} \right) \; \; ,
\eeq

\[ {\bf \vec{\phi}}_{k} = {\rm Re} \phi_{k} \: {\bf \hat{n}}_{k}^{1} + {\rm Im}
\phi_{k} \: {\bf \hat{n}}_{k}^{i} = {\rm Re} (\phi_{k}{\rm e}^{-i\sigma_{k-1}})
\: {\bf \hat{n}}_{k} + {\rm Im} (\phi_{k}{\rm e}^{-i\sigma_{k-1}}) \:{\bf
\hat{b}}_{k}  \; \; ,
\]
and, consequently,
\beq \label{eq:evS}
S_{k,t} = T_{k} S_{k} \; \; ,
\eeq
\[
T_{k} = \frac{i}{2}\left[ \begin{array}{cc}
\gamma_{k} & \delta_{k} \\
\bar{\delta}_{k} & -\gamma_{k} \end{array} \right] \in su(2) \; \; .
\]
The compatibility condition between equations (\ref{eq:S})(\ref{eq:kin_r}) and
(\ref{eq:evS}) specifies the entries $\gamma_{k}$ and $\delta_{k}$ in terms of
the velocity field:
\begin{eqnarray}
\delta_{k} & = & i\zeta^{-2}\phi_{k} - i ( \phi_{k+1} + q_{k} (V_{k} +
V_{k+1})) \nonumber \\
\gamma_{k} & = & W_{k} + \sin(\lambda\Delta) V_{k} \\
W_{k+1}- W_{k} & = & - {\rm Re} \left( i\bar{q}_{k} \left[ \phi_{k+2} -\phi_{k}
+ q_{k+1} (V_{k+1} +V_{k+2}) \right] \right) \nonumber
\end{eqnarray}
and yields the kinematics
\begin{eqnarray} \label{eq:kinAL}
2q_{k,t} & = & -2\cos(\lambda\Delta)\left( \phi_{k+1} + q_{k}V_{k+1} \right) +
(1+|q_{k}|^{2}) \left[ \phi_{k} + \phi_{k+2} + q_{k+1} ( V_{k+1} + V_{k+2})
\right] \nonumber \\
& + & 2iq_{k}W_{k} + q_{k} (q_{k}\bar{\phi}_{k} - \bar{q}_{k}\phi_{k}) \\
& = & -2\cos(\lambda\Delta)\left( \phi_{k+1} + q_{k}V_{k+1} \right) +
{\cal R}\left( \phi_{k+1} + q_{k}V_{k+1} \right) \nonumber \; ,
\end{eqnarray}
\beq
(1-|q_{k}^{2}|) V_{k+1} - (1+|q_{k}|^{2}) V_{k} = q_{k} \bar{\phi}_{k+1} +
\bar{q}_{k} \phi_{k+1} \; \; ,
\eeq
where
\begin{eqnarray}
{\cal R}f_{k} & := & (1+|q_{k}|^{2}) \left( f_{k+1} + f_{k-1} + 2(q_{k+1}E -
q_{k-1} ) (E-1)^{-1} {\rm Re}\frac{\bar{q}_{k}f_{k}}{1+|q_{k}|^{2}} \right)
\nonumber \\
& - & 2iq_{k}(E-1)^{-1}{\rm Im}(\bar{q}_{k+1} f_{k} - \bar{q}_{k} f_{k+1}) \;
\; .
\end{eqnarray}
Substituting the ansatz
\beq
\left[ \begin{array}{c} V_{k} \\ \phi_{k} \end{array} \right] = \sum_{j=0}^{m}
(\cos(\lambda\Delta))^{j} \left[ \begin{array}{c} V_{k}^{(m-j)} \\
\phi_{k}^{(m-j)} \end{array} \right]
\eeq
into equation (\ref{eq:kinAL}) and requiring independence of
$\cos(\lambda\Delta)$, we finally obtain the following class of integrable
dynamics:
\beq
q_{k,t} = h_{0}({\cal R}) (1+|q_{k}|^{2}) (q_{k+1} - q_{k-1}) + h_{1}({\cal
R})(iq_{k}) \; \; ,
\eeq
where $h_{0}$ and $h_{1}$ are arbitrary entire functions with real
coefficients.

We remark that, in the degenerate case of the curve on $S^{2}$ ($q_{k} \in {\bf
R}$) we are forced to choose $h_{1}=0$ and only the first hierarchy survives.

We also remark that equation (\ref{eq:kinAL}) implies the following
interesting connection:
\beq
K_{k}^{(m)} = \phi_{k+1}^{(m)} + q_{k}V_{k+1}^{(m)}
\eeq
between the integrable commuting flows
\beq
K_{k}^{(m)} = {\cal R}^{m-1}(1+|q_{k}|^{2})(q_{k+1} - q_{k-1}) \; \;
{\rm and/or} \; \; K_{k}^{(m)} = {\cal R}^{m}(iq_{k}) \; \; , \; \; m\geq 0
\eeq
and the velocity fields. In the continuous limit this reduces to the result of
\cite{LaPer}.

\bigskip

\noindent The simplest examples are the following:

\medskip

\noindent i) If $h_{0}=1$, $h_{1}=0$, then
\beq
{\bf \vec{v}}_{k} = \frac{\sin(\lambda\Delta)}{\lambda}( {\bf \hat{t}}_{k} -
{\bf \vec{q}}_{k-1} ) = \frac{\sin(\lambda\Delta)}{\lambda}( {\bf \hat{t}}_{k}
 - |q_{k-1}| {\bf \hat{n}}_{k}) \; \; ,
\eeq
\beq
q_{k,t} = (1+|q_{k}|^{2}) (q_{k+1} - q_{k-1}) =: K_{k}^{(1)} \; \; .
\eeq
ii) If $h_{0}(x)=x$, $h_{1}=0$, then
\beq
{\bf \vec{v}}_{k} = \frac{4\sin(\lambda\Delta)}{\lambda} \left\{ \left(
\cos(\lambda\Delta) + \frac{1}{2}(q_{k}\bar{q}_{k-1} + \bar{q}_{k} q_{k-1})
\right) {\bf \hat{t}}_{k} + {\bf \vec{\phi}}_{k} \right\} \; \; ,
\eeq
\[ \phi_{k}= \left( -q_{k-1}\cos(\lambda\Delta) + \frac{1}{2}\left[ (1+
|q_{k-1}|^{2} )(q_{k} - q_{k-2}) - q_{k-1}(q_{k}\bar{q}_{k-1} + \bar{q}_{k}
q_{k-1} ) \right] \right) \]
\beq
q_{k,t} =  K_{k}^{(2)} :=
\eeq
\[
( 1 + |q_{k}|^{2}) \left[ ( 1 + |q_{k+1}|^{2} ) q_{k+2} - ( 1 +|q_{k-1}|^{2} )
q_{k-2}
+  \bar{q}_{k} ( q_{k+1}^{2} - q_{k-1}^{2} ) + q_{k} ( q_{k+1} \bar{q}_{k-1} -
q_{k-1}\bar{q}_{k+1} ) \right] \; \; .
\]
The linear combination
\beq
q_{k,t} = K_{k}^{(2)} - 2 K_{k}^{(1)}
\eeq
reduces, in the continuous limit, to the complex mKdV equation
\beq
\psi_{,t'} = \psi_{,sss} + \frac{3}{2}|\psi|^{2}\psi_{,s} \; \; , \; \;
\psi(s) := \lim_{\Delta\rightarrow 0}\frac{2 q_{k}}{\Delta} = \kappa(s)
{\rm e}^{i\partial_{s}^{-1}\tau} \; \; , \; \; t'= 2\Delta^{3}t \; \; .
\eeq

\medskip

\noindent iii) If $h_{0}=0$ and $h_{1}(x)=x$, we obtain
\beq
{\bf \vec{v}}_{k} = \frac{2\sin(\lambda\Delta)}{\lambda} \: ({\bf \vec{i
q}}_{k-1})  = \frac{2\sin(\lambda\Delta)}{\lambda} \: \tan(\frac{\vf_{k-1}}{2})
{\bf \hat{b}}_{k} \; \; ,
\eeq
\beq \label{eq:dNLS}
q_{k,t} = i(1+ |q_{k}|^{2}) (q_{k+1} + q_{k-1}) \; \; .
\eeq
The following combination of equation (\ref{eq:dNLS}) with the "zero order
flow" $q_{k,t}= i q_{k}$ :
\beq \label{eq:dNLS2}
q_{k,t} = i\left[ q_{k+1} - 2q_{k} + q_{k-1} + |q_{k}|^{2}(q_{k+1} +
q_{k-1}) \right]
\eeq
reduces \cite{AL}, in the continuous limit, to the NLS equation
\beq
i\psi_{,t} = \psi_{,ss} + \frac{1}{2}|\psi|^{2}\psi \; \; , \; \;
t'=-\Delta^{2} t \; \; .
\eeq
which describes the motion of a vortex filament in the
localized induction approximation \cite{Hasimoto}\cite{Batch}. We remark that,
in this approximation, the velocity field which governs the motion of the
vortex depends on its curvature $\kappa$ through the relation
\beq
{\bf \vec{r}}_{,t} = \kappa {\bf \hat{b}} \; \; ;
\eeq
therefore since equation (\ref{eq:dNLS2}) has in ${\bf R}^{3}$ a velocity field
of the same type:
\beq
{\bf \vec{r}}_{k,t} = 2 \Delta \tan(\frac{\vf_{k-1}}{2}) {\bf \hat{b}}_{k}\;
\; ,
\eeq
we expect it to be a good candidate for describing the motion of a discrete
vortex in the same approximation.

\section{Acknowledgements}
 We would like to thank A. B. Shabat who encouraged this research, A. Sym
for pointing out ref. \cite{Aminov} and D. Levi and M. Bruschi for useful
discussions.

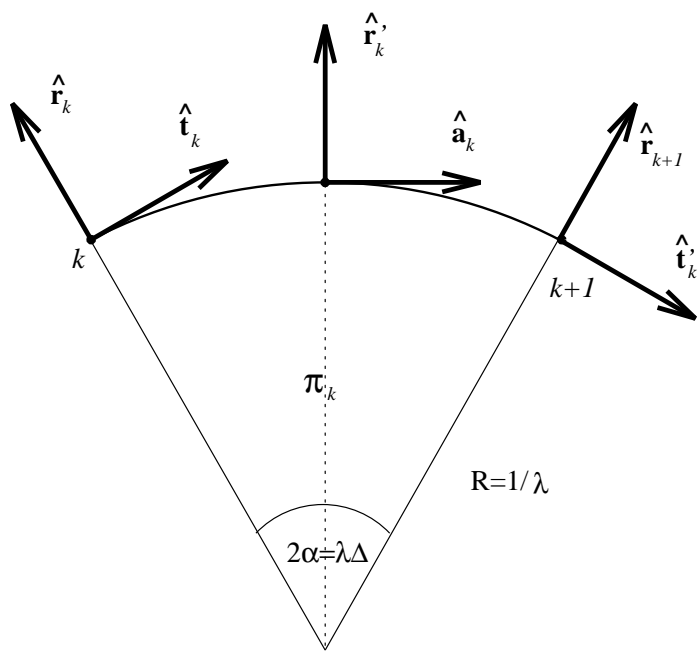
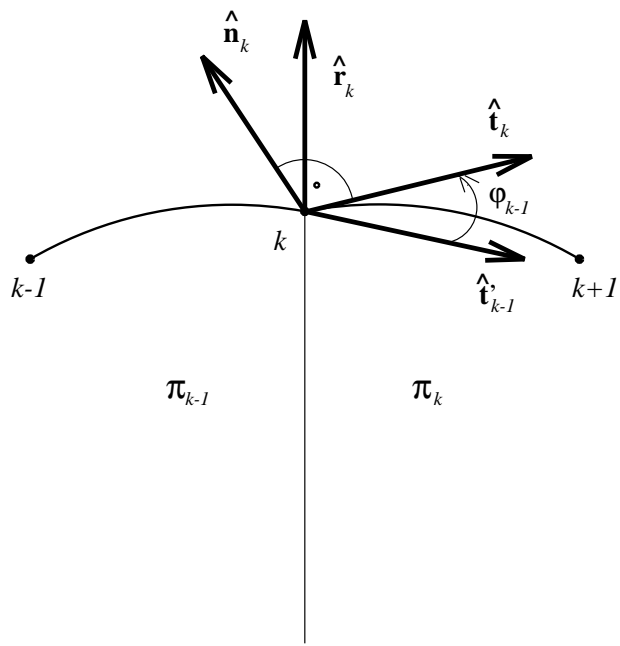
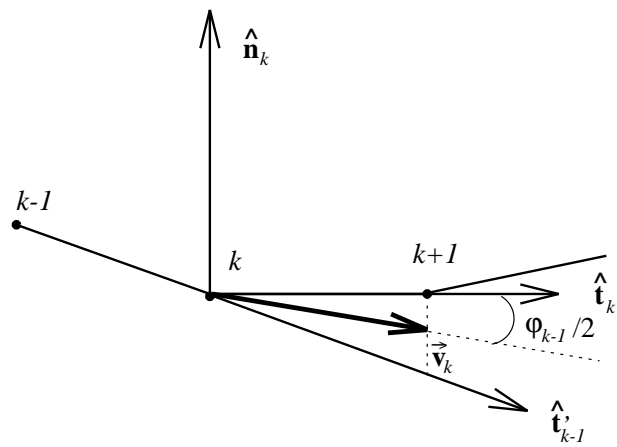
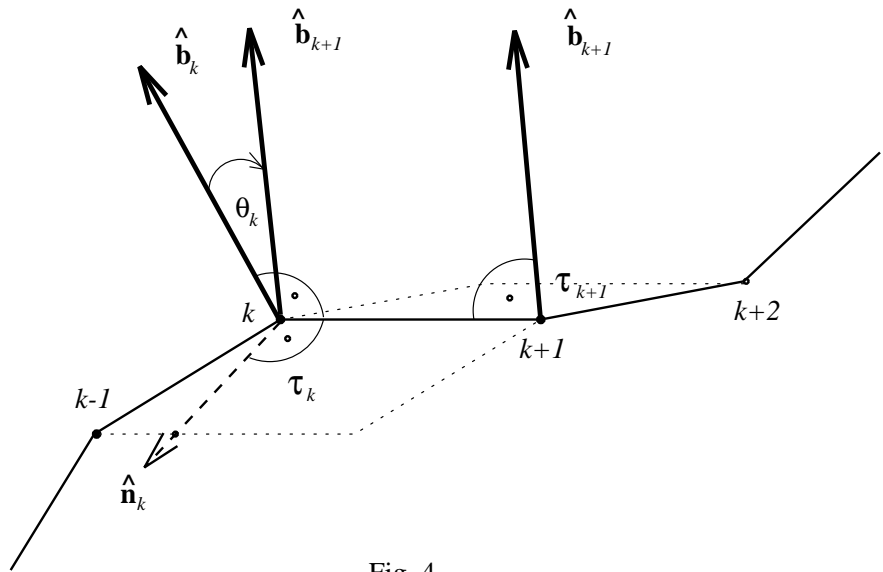
\end{document}